\documentstyle [12pt,epsf] {article}

\newcommand{\tr}{{\rm Tr}}
\newcommand{\re}{{\rm Re}}

\newcommand{\retr}{\re\,\tr}
\newcommand{\err}[1]{{{$(\makebox[1.3em][r]{#1})$}}}
\newcommand{\formul}{$y=\sigma^{1/2}/\Lambda_{\overline{MS}}$~}

\begin{document}
\begin{titlepage}
\rightline{WUB 93--37}
\vskip 3cm
\begin{center}
\LARGE
{\bf The Running Coupling From Lattice QCD\footnote{Talk presented at
the XVI Workshop on High Energy Physics, Sept.~14--17 1993, Protvino, Russia.}
}
\end{center}
\vskip 1cm
\centerline{\bf Gunnar S.~Bali\footnote{E-mail:
bali@wpts0.physik.uni-wuppertal.de}}
\centerline{Fachbereich Physik, Bergische Universit\"at,
             Gesamthochschule Wuppertal}
\centerline{Gau\ss{}stra\ss{}e 20, 42097 Wuppertal, Germany}
\vskip 1cm
\centerline{\bf ABSTRACT}
\sl
\small
A recent lattice calculation of the QCD running coupling is
presented. The coupling is extracted from the force between two static
quarks in the framework of the valence quark approximation.
A value of the $\Lambda$-parameter for zero quark flavours is determined:
$\Lambda_{\overline{M\!S}}^{(0)}=0.630(38)\sqrt{\sigma}
=293(18)^{+25}_{-63}$~MeV.
The first error is statistical, the second stems from the overall scale
uncertainty in the string tension $\sigma$. Combining this value with
results from full QCD lattice simulations, we end up with the estimate
$\Lambda^{(4)}_{\overline{MS}}=129(8)^{+43}_{-60}$~MeV or
$\alpha_{\overline{MS}}(m_Z)=0.102^{+06}_{-11}$.
\end{titlepage}

\normalsize
\section{Introduction}
QCD, the theory of strong interactions, contains $8$ free parameters,
namely the quark masses, a
(${\cal CP}$-violating) vacuum angle $\Theta$ which is consistent
with {\em zero}, and
the running coupling parameter, that can only be fixed by
experiment. For an empirical test of QCD, the following four steps have to
be performed:
\begin{enumerate}
\item{measurement of $8$ independent observables (e.g.\ particle masses),}
\item{fixing the free parameters by use of QCD,}
\item{predicting a value for another observable,}
\item{comparison with experiment.}
\end{enumerate}
It turns out that the second step is highly non trivial. So,
most of this talk will deal with this task. I will use experimental
data on the mass of the $\rho$-meson $m_{\rho}$,
and the string tension $\sigma$,
obtained from the quarkonia spectra, in order to predict
the running of the coupling from QCD, which can then be compared to
experiment. Since the precision of the lattice determination of $\alpha_S$
can compete with the experimental
{\em status quo}, the result can be
used as an input for phenomenological calculations.

There are different ways to define a running coupling.
Either one can state a value for $\alpha_S(\mu)$ at some fixed scale
$\mu$ (like the mass of the $Z^0$ boson $m_Z$), the momentum $\mu$, at which
the coupling has a given strength, or a $\Lambda$-parameter from the
functional dependence $\alpha(q)=f(q/\Lambda)$, which can be obtained
in perturbation theory, for high momenta transfer $q$. Here, I take a
low energy scale $\mu$, and calculate $\alpha_S(N\times\mu)$ on the
lattice with $N$ being a large
number, in order to make contact with energies at
which perturbation theory becomes
valid. It is in this region where one can translate one
scheme into another.

The only known method which
allows to treat QCD problems
\begin{itemize}
\item{nonperturbatively,}
\item{quantitatively,}
\item{from first principles, and}
\item{with control over possible systematic errors,}
\end{itemize}
is Lattice Gauge Theory. Unfortunately, it is a stochastical
(non-exact) method, and thus statistical errors are introduced. Also,
computational resources are a serious limitation.
Since speed of supercomputers evolves
according to an exponential scaling law
in time, the latter problem seems to vanish. In the following, I will
give a brief methodological introduction into lattice gauge theory,
and comment on the advantages and caveats. This is followed by a
section about how to determine the physical spacing of a given
lattice, a section about improved perturbation theory on the lattice,
and a section about our actual calculation of $\alpha_S$ from the
interquark force, and how to relate the result obtained to schemes
which are more commonly used in perturbative QCD. At the end an
outlook is given.

\section{Lattice Gauge Theory}

The first step in introducing the lattice formulation of a gauge
theory is a rotation from Minkowski to Euclidean space ($t\rightarrow
it$). Because of this rotation, information about real time dynamics is
lost but static quantities like the mass spectrum remain unharmed.
Then, space-time is approximated by a four dimensional hypercubic
lattice with spacing $a$ and extent $La$. $\pi/a$ serves as the
required ultraviolett cutoff.
After this is done, calculations of quantities of interest on the
lattice are performed. Since a stochastic (Monte-Carlo) method is used,
one also describes this procedure with the term ``measurements are
taken''. Finally, the results are translated back to the
continuum. For this step, various extrapolations become
necessary. The limit $a\rightarrow 0$ (that corresponds to the limit
$g^2\rightarrow 0$ in asymptotically free theories)
can be calculated by use of ordinary
perturbation theory in the coupling parameter $g^2$, once the
perturbative region is reached. The infinite volume limit
$La\rightarrow\infty$ can be reached by use of Finite Size Scaling
methods which are well established in statistical mechanics, and the
extrapolation to small quark masses is controlled by chiral
perturbation theory.

In the following, I will briefly describe some basic notations, used
in lattice QCD. The Wick-rotated action $S_{QCD}=S_g+S_f$
contains a purely gluonic part, $S_g$, which
can be written as
\begin{equation}
S_g=-\frac{1}{4}\int\!d^4\!x\,F_{\mu\nu}^aF_{\mu\nu}^a=
-\beta\underbrace{\sum_{n,\mu >\nu}\left(1-\frac{1}{N}\retr\{
U_{\mu\nu}(n)\}\right)}_{S_W}+{\cal O}(a^2)
\end{equation}
with $\beta=2N/g^2$, and the number of colours $N=3$.
$g^2$ is the bare lattice coupling. The discretized version of the
gluonic action,
$S_W$, is the so-called Wilson action. The continuum four-vector
$x$ is mapped to the discrete vector $n=x/a$. Instead of non-compact
gauge fields, $A_{\mu}(x)$, $SU(3)$ matrices are introduced:
\begin{equation}
U_{\mu}(n)
={\cal P}\left[\exp\left(ig\int_{na}^{n(a+\hat{\mu})}\!\!\!dx\,A_{\mu}(x)
\right)\right]\quad.
\end{equation}
In terms of these, the gauge symmetry is
easier to preserve since products of such ``link-variables'' along a
closed path, e.g.~the Wilson action, are automatically gauge invariant. The
lattice version of the Maxwell field strength tensor, then, is the
ordered product of link variables around an elementary square (plaquette)
$U_{\mu\nu}(n)=\exp\left(iga^2\left(F_{\mu\nu}(an)+{\cal
O}(a)\right)\right)$. In the following it will also be referred to as
$U_{\Box}$.

Expectation values of observables $O$ are calculated in the path integral
approach as their mean value over all possible field configurations in the
finite box
\begin{equation}
\langle
O\rangle=\frac{1}{Z}\int\!\left(\prod_{n,\mu}dU_{\mu}(n)\right)\,
O\,e^{-\beta S_W}\quad,\quad Z=\int\!\left(\prod_{n,\mu}dU_{\mu}(n)\right)\,
e^{-\beta S_W}\quad.
\end{equation}
In order to solve this high dimensional integral, $N$
independent field configurations $C_i=\{U_{\mu}^{(i)}(n)\}$ are
generated according to the measure
$dP(C_i)=\left(\prod_{n,\mu}dU_{\mu}^{(i)}(n)\right)\,e^{-\beta S_W(C_i)}$.
The expectation value is calculated on a ``representative
ensemble''
\begin{equation}
\langle O\rangle = \lim_{N\rightarrow\infty}\frac{1}{N}\sum_{i=1}^N
O(C_i)\quad.
\end{equation}
This gives rise to a statistical error, $\Delta O$, which reduces as $N$ is
increased: $\Delta O\propto 1/\sqrt{N}$.

In the present calculation, the valence quark (quenched)
approximation ($S_f=$~const.) is used, in which vacuum
polarization effects, caused by sea quarks, are
neglected. In the language of perturbation theory this means that fermionic
loops are dropped while gluonic diagrams are taken to all orders. The
reason for this approximation are limited computational resources.
However, all mass ratios from real world experiments are
reproduced within an error of $6\%$~\cite{weingarten}. So, if full QCD
with its dynamical fermions
describes nature, discrepancies between full and quenched results are
likely to coincide within $10\%$ or so. This claim is also supported by
the success of simple models like the naive quark model.

\section{How to set the mass scale}

In quenched lattice QCD, the inverse bare coupling $\beta$ is the only
free external parameter. So, the value of $\beta$ determines the masss scale
(and lattice spacing $a$): One measures a physical mass
$M$ in lattice units, and fixes the lattice spacing by the
corresponding experimental value $m_{phys}$:
$a(\beta)=m_{phys}/M(\beta)$. On repeating this procedure for
various $\beta$ one can determine the dependence of the spacing on the
bare coupling $a(g^2)$, or, by inverting this relation, extract a
running coupling $g^2(a)$. As soon as a behaviour according to
perturbation theory is observed (asymptotic scaling), this running
lattice coupling can be translated into any other renormalization
scheme perturbatively.

The present simulations have been performed at the $\beta$ values
$\beta=$ 5.5, 5.6, 5.7, 5.8, 5.9, 6.0, 6.2, 6.4 and 6.8. $L=16$ was
chosen for the first five values, and $L=32$ for the latter four.
Below $\beta=5.5$ ($a\approx 0.25$ fm) agreement with strong coupling
expectations is found. Between 5.5 and 5.7 ($a\approx 0.18$ fm), as the
(largest) correlation length in lattice units increases up to a few lattice
spacings, a transition from discrete hypercubic symmetry to
continuum rotational symmetry
$\left(\sum_i\left|R_i\right|\longrightarrow\sqrt{\sum_i R_i^2}\right)$
is observed on the scale of a few $a$. Between 5.7
and 6.0 ($a\approx 0.1$ fm) ratios of physical masses, extracted from
the lattice, remain constant within about $10\%$. From 6.0 up to 6.8
($a\approx 0.03$~fm), these ratios remain constant within
statistical accuracy of contemporary lattice simulations.
This is referred to as scaling in contrast
to asymptotic scaling that does not set in for $\beta<{\cal O}(10)$
at least.

In order to fix the physical mass scale, we use the string tension
$\sigma=K/a^2$ which is defined to be the slope of the potential
between heavy quark sources $V(R)$ at large $q\bar{q}$-separations
\begin{equation}
K=-\lim_{R\rightarrow\infty} F(R)\quad ,\qquad F(R)=-\frac{\partial
V(R)}{\partial R}\quad .
\end{equation}
The obvious advantages in extracting the mass scale from the potential
can be summarized as follows:
\begin{itemize}
\item The measurement of $V(R)$ consumes little computer time since no
matrices have to be inverted for calculation of propagators.
\item Finite size effects are negligible for $La>1$ fm~\cite{wir,peran}.
\item No extrapolations in the quark masses are needed.
\item The string tension is the most accurately measured dimensionful
quantity on the lattice~\cite{wir2,wir3}.
\end{itemize}
In the following I will discuss, which physical scale corresponds to
$\sqrt{\sigma}$ and how to
define it reasonably on a lattice with finite extent.

Apart from investigations of Regge trajectories which suggest
values $\sqrt{\sigma}\approx$ 400--450 MeV,
potential related information can be gained
from the Charmonium and Bottomium spectra by integration of the
Schr\"odinger equation~\cite{eicht1,quigg,eicht2}.
It is practically only the potential at distances $r$ between about 0.2~fm
and 1~fm that determines these quarkonia spectra, since the
corresponding wavefunctions almost vanish for smaller or larger
$q\bar{q}$ separations.
This is a lucky incidence because for smaller $r$
relativistic corrections to the Schr\"odinger equation
would have to be taken into account. For larger
separations ($r>1$~fm),
differences between the quenched potential (without sea
quarks) and the real
QCD potential are expected due to creation of a
$q\bar{q}$ pair from the vacuum (string breaking).
The phenomenological potential is well
described by the Cornell parametrization
\begin{equation}
\label{cornell}
V(R)=V_0-\frac{e}{R}+KR\quad.
\end{equation}
Before discovery of the $\Upsilon'$, typical fit parameters have been
$\sqrt{\sigma}=$ 455 MeV, and $e=0.25$~\cite{eicht1}. After inclusion
of the $\Upsilon$ mass splitting, these parameters moved rapidly to
values like $\sqrt{\sigma}$=412 MeV, $e=0.51$~\cite{quigg} or
$\sqrt{\sigma}$=427 MeV, $e=0.52$~\cite{eicht2}. A careful inspection
shows that the $e\approx 0.25$ and $e\approx 0.5$ parametrizations
of the potential differ only weakly within the accessible range
(0.2 to 1~fm). This is related to the fact
that the fit parameters are highly correlated.

\begin{figure}[thb]
\begin{center}
\leavevmode
\epsfxsize=400pt
\epsfbox[0 160 600 560]{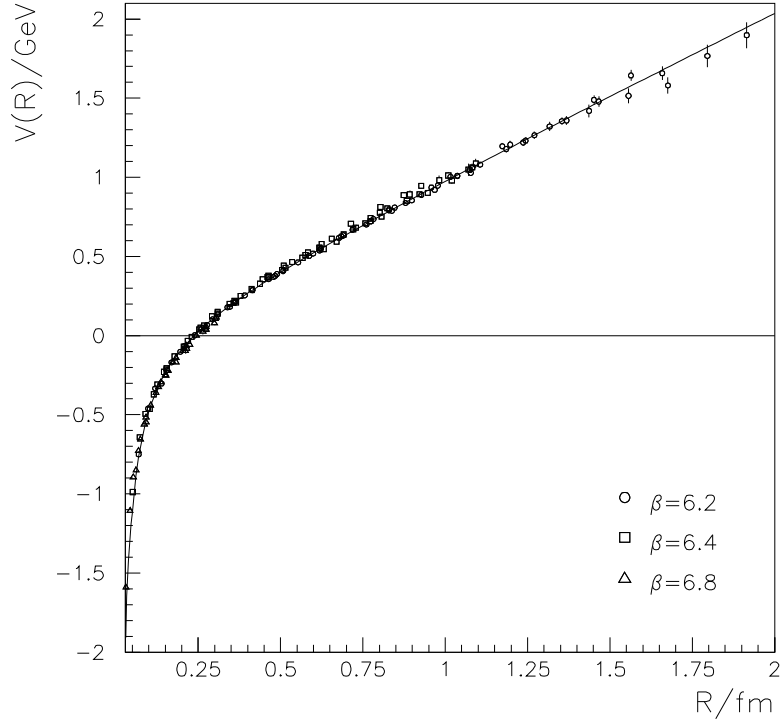}
\end{center}
\caption{\em The (quenched) QCD potential $V(R)$, measured at
$\beta=6.2$, $\beta=6.4$, and $\beta=6.8$ as a function of
$r$ in physical units. The mass scale is extracted from the string tension.}
\label{fig1}
\end{figure}

\begin{figure}[thb]
\begin{center}
\leavevmode
\epsfxsize=290pt\epsfbox[50 160 650 560]{fig2a.eps}\epsfxsize=290pt\epsfbox[210
160 810 560]{fig2b.eps}
\end{center}
\caption{\em The fit parameter $e$ versus $K^{1/2}$
(Eq.~\protect\ref{cornell}), and the redefined string tension,
calculated from $R_0^{-1}$ (Eq.~\protect\ref{r0}). The distribution of
$1000$ bootstrap samples is shown. The higher the density of the
points, the larger is the corresponding probability.}
\label{fig2}
\end{figure}

We have attempted to fit our lattice
potential (Fig.~\ref{fig1}) for a fixed physical
cutoff $Ra\geq 0.3$ fm\footnote{For $\beta=5.5$ and $\beta=5.6$,
$Ra\geq 0.5$~fm, $Ra\geq 0.4$~fm have been chosen, respectively.},
and different $\beta$ to the Cornell form. In
Fig.~\ref{fig2}a we display the scatter of fit parameters among different
bootstrap samples in the $K$-$e$-plane where we observe a
correlation, similar to the ambiguity
in parametrizing the quarkonium potential. To circumvent this problem,
Sommer has invented a more stable scale from the
potential~\cite{sommer} that is also well-defined in the
screened case of full QCD. The idea is to calculate the dimensionless
quantity $X(R_0)=-F(R_0)R_0^2$ from the interquark force. Since the
potential models are most sensitive to the spectrum at a distance
of about 0.5 fm, a scale $r_0=R_0a\approx 0.5$ fm is defined
by imposing the condition
$X(R_0)=1.65$.
Using the fit parameters of Ref.~\cite{eicht1} one ends up with
$r_0^{-1}= 384$ MeV. Ref.~\cite{quigg} yields $r_0^{-1}= 386$ MeV
and Ref.~\cite{eicht2} $r_0^{-1}= 402$ MeV. Another set of fit
parameters, taken from Ref.~\cite{eicht2}, leads to $r_0^{-1}= 416$ MeV.
We conclude with the value
$r_0^{-1} = (400\pm 15)$~MeV.

\begin{table}
\begin{center}
\begin{tabular}{|c|c|c|}
\hline
$\beta $&$S_{\Box}$&$\sqrt{\sigma_r}a$
\\\hline
5.5&0.503196\err{18}&0.5805\err{ 84}\\
5.6&0.475495\err{27}&0.5092\err{136}\\
5.7&0.450805\err{25}&0.4069\err{ 78}\\
5.8&0.432349\err{21}&0.3206\err{ 41}\\
5.9&0.418164\err{15}&0.2518\err{ 62}\\
6.0&0.406318\err{ 5}&0.2204\err{ 17}\\
6.2&0.386369\err{ 3}&0.1582\err{  9}\\
6.4&0.369364\err{ 2}&0.1179\err{  9}\\
6.8&0.340782\err{ 4}&0.0698\err{ 18}\\
\hline
\end{tabular}
\end{center}
\caption{\em The average plaquette action, and the redefined string
tension in lattice units.}
\label{tab1}
\end{table}

Deviating from Sommer's prescription, we fit the potential to the
Cornell ansatz and calculate
\begin{equation}
\label{r0}
R_0^{-1}=\sqrt{\frac{K}{1.65-e}}
\end{equation}
from the fitted parameters. All fits yield effective Coulomb
coefficients consistent with the value $e=0.296$, determined from the
$\beta=6.2$ data that carries the smallest errors.
For convenience, we rescale the value of $r_0^{-1}$ by the factor
$\sqrt{1.65-0.296}=1.164$ in order to obtain a string tension
$\sigma_r=1.164r_0^{-1}$. A bootstrap scatter plot in the
$K$-$e$-plane is displayed in Fig.~\ref{fig2}b. It can be seen that this
definition of a string tension is insensitive towards the value of the
Coulomb coefficient $e$. The string tension results are summarized in
Tab.~\ref{tab1}.
A detailed description of the data evaluation and fitting procedures
will be given elsewhere~\cite{wir4}. The physical value that
corresponds to the redefined
string tension is
\begin{equation}
\label{string-v}
\sigma_r=465\pm 18\mbox{~MeV}\quad .
\end{equation}

While ratios of pure glue quantities like the $0^{++}$ or $2^{++}$
glueball masses in units of the string tension exhibit good scaling
behaviour above $\beta=5.7$~\cite{ukqcd}, the situation concerning
observables with valence quark content is less satisfactory at present
values of $\beta$. In order to
take these effects into account we have extrapolated our string tension
values to $\beta=5.93$ and $\beta=6.17$, using the $E$-scheme
that will be described below, and calculated the ratio of the $\rho$
mass, taken from Ref.~\cite{weingarten}, over the string tension at
these $\beta$-values and at $\beta=5.7$. After this, linear and quadratic fits
on the $a$ dependence of this ratio have been performed, yielding a
continuum value of $2>m_{\rho}/\sqrt{\sigma}>1.7$. This corresponds to
385 MeV $<\sqrt{\sigma}<$ 455 MeV. The difference between this range
and the number from the potential analysis (Eq.~\ref{string-v}) might be
caused by both, scaling violations still present above $\beta=6.0$,
and quenching. If we take this uncertainty into account and
allow for an additional $5\%$ quenching error, we
end up with the conservative scale estimate
\begin{equation}
\label{sca}
\sigma_r=465^{+40}_{-100}\mbox{~MeV}\quad .
\end{equation}
By full QCD simulations the huge systematic error can probably be
reduced to about $5\%$ or even less. In the following, $\sigma_r$ will
be abbreviated as $\sigma$, and $\sigma_r a^2$ as $K$.

\section{Asymptotic scaling and redefined couplings}

\begin{figure}[htb]
\begin{center}
\leavevmode
\epsfxsize=400pt
\epsfbox[0 160 600 560]{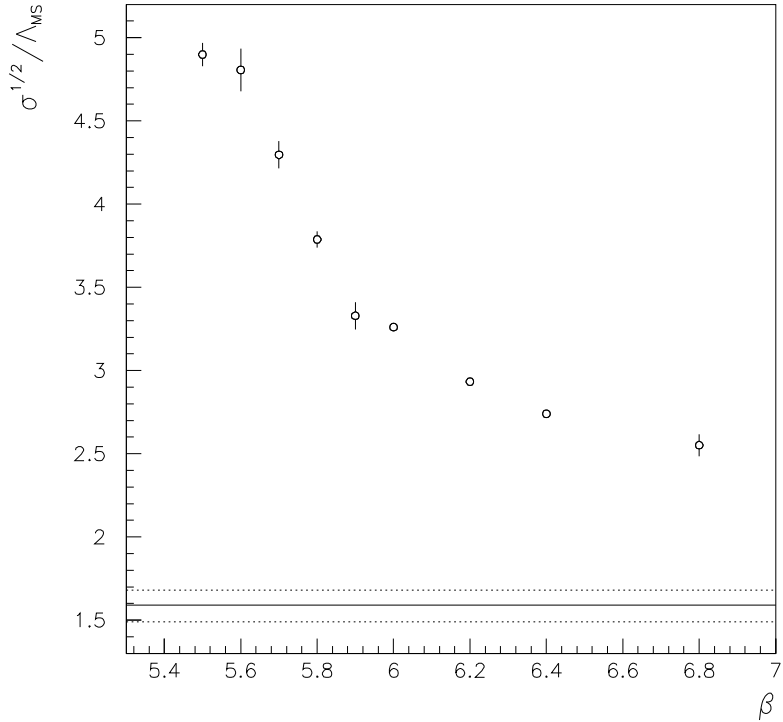}
\end{center}
\caption{\em The ratio \formul in the
two-loop approximation, calculated by
use of the bare coupling $g^2=6/\beta$.
The error band denotes the value $y=1.59\pm 0.10$,
extracted from the running of the coupling $\alpha_{q\bar q}$ from the
interquark force.}
\label{fig3}
\end{figure}

The QCD $\beta$-function can be expanded perturbatively
\begin{equation}
-\frac{\partial g^2}{\partial \ln a}=-b_0g^3-b_1g^5+\cdots\quad.
\end{equation}
Higher order terms are regularization scheme and gauge dependent. The
first two coefficients are given by:
\begin{equation}
b_0=\frac{11N-2n_f}{3(16\pi^2)}\quad,\quad
b_1=\frac{34N^3-(13N^2-3)n_f}{3N(16\pi^2)^2}\quad.
\end{equation}
$n_f$ denotes the number of quark flavours, $N$ the number of colours.
Integrating this $\beta$-function leads to
\begin{equation}
\label{twolo}
a\Lambda_L=f(g^2):=
e^{-1/2b_0g^2}\left(b_0g^2\right)^{-b_1/2b_0^2}\quad,
\end{equation}
where $\Lambda_L$ is an integration constant, not predicted by the theory.
For sufficiently small $g^2$, the ratio
$\sqrt{\sigma}/\Lambda_L=\sqrt{K}/a\Lambda_L=\sqrt{K}/f(g^2)$ is
expected to remain constant. This perturbative behaviour is
called ``asymptotic scaling''. As can be seen from Fig.~\ref{fig3} there are
large deviations from the asymptotic expression for $\beta<7$. On the
other hand it is the $\beta$ region $\beta>6$ where good scaling
between physical quantities like mass ratios is observed. So it might
be that just the bare coupling is an ill behaved expansion
parameter~\cite{mack1}. The situation might be improved if one chooses a more
appropriate renormalisation scheme by use of a more ``physical''
coupling parameter.

Several suggestions have been made in the past for ``better''
couplings. I will review three such choices which have been proven
to be successful, at least in the $\beta$ range accessible to
present day computer simulations. All are based on the idea of mean
field improvement and were stimulated by Parisi~\cite{parisi}. The Fermilab
group proposed an improved coupling
$g^2_{\mbox{\scriptsize FNAL}}$~\cite{mack,mack2}
which corresponds to
the continuum modified minimal subtraction ($\overline{M\!S}$)
scheme~\cite{msbar} in
the limit $a\rightarrow 0$. A perturbative calculation~\cite{Hasenfratz} shows
\begin{equation}
\frac{1}{g^2_{\overline{M\!S}}(\mu)}=\frac{1}{g^2(\mu)}
\underbrace{-d+0.008204N+0.00278n_f}_{\Delta}+\cdots\quad,
\quad d=\frac{N^2-1}{8N}\quad.
\end{equation}
The large correction coefficient $d$, caused by tadpole diagrams, leads to
the relation $\Lambda_{\overline{M\!S}}=\pi e^{-\Delta/2b_0}\Lambda_L=
28.81\Lambda_L$ for $N=3$,
$n_f=0$.
On the lattice,
the average plaquette action $\langle S_{\Box}\rangle=\langle
1-U_{\Box}\rangle$ can be expanded in powers of
$g^2$~\cite{digiac,oneloop,pana}
\begin{equation}
\label{plaq-exp}
\langle S_{\Box}\rangle =c_1g^2+c_2g^4+c_3g^6+\cdots\quad.
\end{equation}
The fields $A_{\mu}(x)$ are expected to
fluctuate around {\em zero}. So the mean field action is expected to
be {\em zero}.
Since the coefficient $c_1$, that causes the deviations from this
expectation, turns out to
equal the large number $d$, that has been responsible for
the huge difference between the two $\Lambda$-parameters, the idea
is to substitute $d=c_1$ by $\langle
S_{\Box}\rangle/g^2$. This leads to the redefined coupling
\begin{equation}
\frac{1}{g^2_{\mbox{\scriptsize FNAL}}(\pi/a)}=\frac{\langle
U_{\Box}\rangle}{g^2(\pi/a)}
+0.008204N+0.00278n_f
\end{equation}
with $\Lambda_{\mbox{\scriptsize FNAL}}=\Lambda_{\overline{M\!S}}$.
In Ref.~\cite{mack2} a different scheme is introduced, in which a coupling
$g^2_V$ is defined from logarithms of small Wilson loops.

Another empirical concept is the $E$-scheme, proposed in
Ref.~\cite{karsch}. The starting point
is the observation that large deviations from
asymptotic scaling exist as well as many quantities like the average
plaquette are not well described by perturbation theory. The idea, now,
is that the contributions,
polluting the perturbative expansion of the plaquette, might be similar
to those spoiling asymptotic scaling. By defining a coupling from
inverting Eq.~\ref{plaq-exp}, one forces the plaquette to behave
according to perturbation theory, and hopes to find asymptotic scaling
for dimensionful quantities as well. The $E$ coupling is defined by
$g_E^2=\langle S_{\Box}\rangle/c_1$ with
$\Lambda_E=e^{c_2/\left(2c_1b_0\right)}\Lambda_L=2.0756\Lambda_L$ for $N=3$,
$n_f=0$. Another suggestion~\cite{wir2,wir1} to define a $g^2_{E2}$
by inverting Eq.~\ref{plaq-exp} after the second order. Then
$\Lambda_{E2}=\Lambda_L$.

\begin{figure}[htb]
\begin{center}
\leavevmode
\epsfxsize=400pt
\epsfbox[0 160 600 560]{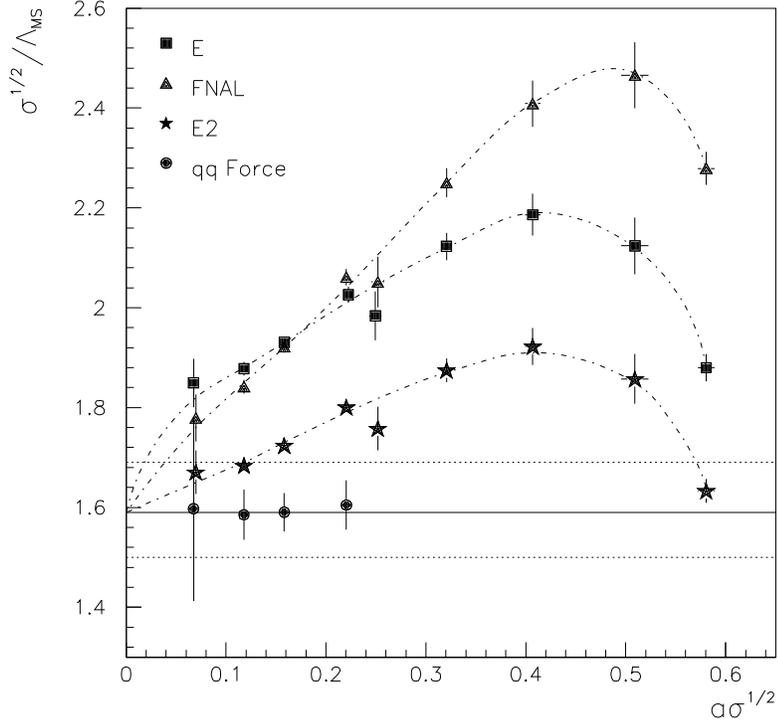}
\end{center}
\caption{\em The ratio \formul
in the
two-loop approximation, calculated by
use of the improved couplings versus the lattice spacing.
The error band denotes the value $y=1.59\pm 0.10$,
extracted from the running of the coupling $\alpha_{q\bar q}$ from the
interquark force (circles). The curves are just drawn to guide the eye.}
\label{fig4}
\end{figure}

In Fig.~\ref{fig4} the effective
$\Lambda_{\overline{M\!S}}^{-1}$, obtained by use of
the two-loop $\beta$-function is plotted in units of the string
tension versus the lattice spacing $a$ for
the above three effective couplings. We find that the behaviour is
improved in all cases, though the $E$-schemes appear to be more stable
than the FNAL-$\overline{M\!S}$-scheme. The differences between the
values, obtained from the effective schemes reflect
still present deviations from asymptotic scaling. One might attempt to
extrapolate $\sqrt{\sigma}/\Lambda_{\overline{M\!S}}$ to $a=0$. If the
corrections to the two-loop formula~Eq.~\ref{twolo}
can be resummed into an effective
third order perturbative coefficient one might attempt to fit the data
to the parametrization
\begin{equation}
\label{fitcurve}
y\left(a(g^2)\right)=y(0)\left(1-{\cal O}(g^2)\right)=
y(0)\left(1+\frac{C}{\ln
\left(y(0)/(a\sqrt{\sigma})\right)}\right)\quad,\quad
y(a):=\frac{\sqrt{\sigma}}{\Lambda_{\overline{M\!S}}(a)}
\end{equation}
where $y(0)$ and $C$ are fit parameters.

Another possibility that works equally well is to argue that the deviations
are of nonperturbative origin and might be parametrized by a
polynomial in the lattice spacing $a$. For $5.7<\beta\leq 6.8$ the
data is well described by a linear fit, although from
Eq.~\ref{fitcurve} it is clear that the behaviour might change rapidly
as one approaches the continuum limit.
As can be imagined from Fig.~\ref{fig4},
with the described variety of fitting
opportunities the extrapolated ratio might be anywhere within the
range $1.4<\sqrt{\sigma}/\Lambda_{\overline{M\!S}}(0)<1.8$.

\section{Direct ``measurement'' of $\alpha_S$}

So far we have inserted a coupling $g^2$ (or calculated an effective
$g^2$ at a fixed $\beta$), and measured the
corresponding scale $a(g^2)$.
The effective couplings are also calculated from the
plaquette, i.e.\ on a characteristic length scale $a$.
This might cause nonperturbative deviations
from asymptotic scaling that depend on the chosen discretization of
the action. In order to reduce
these ``discretization errors'', one should determine a
coupling from observables with a
typical scale of a few lattice spacings, instead.
So far there have been three such approaches.
\begin{enumerate}
\item Extracting couplings from logarithms of small Wilson loops by
tadpole-improved perturbation theory~\cite{elkhadra}.
\item Introducing a volume dependent coupling~\cite{luesch} from the
field response to the boundary conditions and
exploiting finite size matching techniques on small volumes~\cite{luescher}.
\item Calculating $\alpha$ from a ``physical'' quantity like the
interquark force~\cite{chris,wir2,wir1}.
\end{enumerate}
All these methods have their specific merits and disadvantages.
Nonetheless, the results are fairly consistent with each other.
In the first approach, it is not {\em a priory} obvious, what
effective momentum
corresponds to a Wilson loop with a certain extent~\cite{mack2}. Also,
one still encounters deviations from asymptotic behaviour that have to
be examined carefully by simulating at various $\beta$. An obvious
advantage of this method is that Wilson loops are amongst the most simple
quantities that can be calculated on the lattice. The second
approach is a bit awkward since the boundary conditions have to be
fixed. Thus, the generated configurations can only be used for the single
purpose of calculating the coupling. While in the other two methods,
$\alpha(\mu)$ can be computed from one
simulation. Since the
coupling is defined in a technical way (that allows for an easier
perturbative expansion around the fixed background), the
coefficient, connecting this scheme to the continuum motivated
$\overline{M\!S}$ scheme,
is large, and two-loop perturbative behaviour is only found for rather
high energies. At present, it is not clear how
this method can be
generalized to full QCD with its dynamical fermions.
The advantage is that the energy scale can be doubled
with much less effort than when using a single-lattice
method like approach 3, which I am going to discuss in
the following.

The intuitive picture behind the concept of a coupling is that of
local interactions. These are only realized at large energies where
perturbation theory becomes a valid approximation. In principle, any
dimensionless quantity that depends on a mass scale can be
called a coupling. In the case of the quenched approximation to QCD,
only one such mass scale exists.
The $\beta$-function, i.e.\ the logarithmic derivative of the coupling
with respect to the mass, is universal up to two loops perturbation theory.
Just the renormalization point differs from scheme to
scheme. The schemes can be translated
into each other by perturbation theory at large energies (small couplings).
In the lattice formulation
it is possible to define various nonperturbative gauge invariant
couplings which can only be connected with each other in this high
energy region. So one is
forced to use a small lattice spacing, in order to make the results
applicable in the framework of perturbation theory. A volume
$(La)^3 >1$~fm$^3$ is needed
to minimize finite size effects on the $q\bar{q}$-potential~\cite{wir,peran}.
Since the computer memory of
our machine (an 8K CM-2) is limited to $256$ MBytes $L\leq 32$
has to be chosen. These two constraints (the physical and the
technical) set the limit
$a^{-1}<$~6~GeV.
Only potential values from $R\geq\sqrt{2}$ are included
into the analysis of the interquark force,
in order to minimize discretization errors.
As a check whether these are eliminated,
simulations at different spacings $a$ are needed. Fortunately,
clearly perturbative behaviour is found for distances as small as
$1$ GeV$^{-1}$, i.e.\ over a scale factor 3--4 within the
accessible energy range.

\begin{figure}[htb]
\begin{center}
\leavevmode
\epsfxsize=400pt
\epsfbox[0 160 600 560]{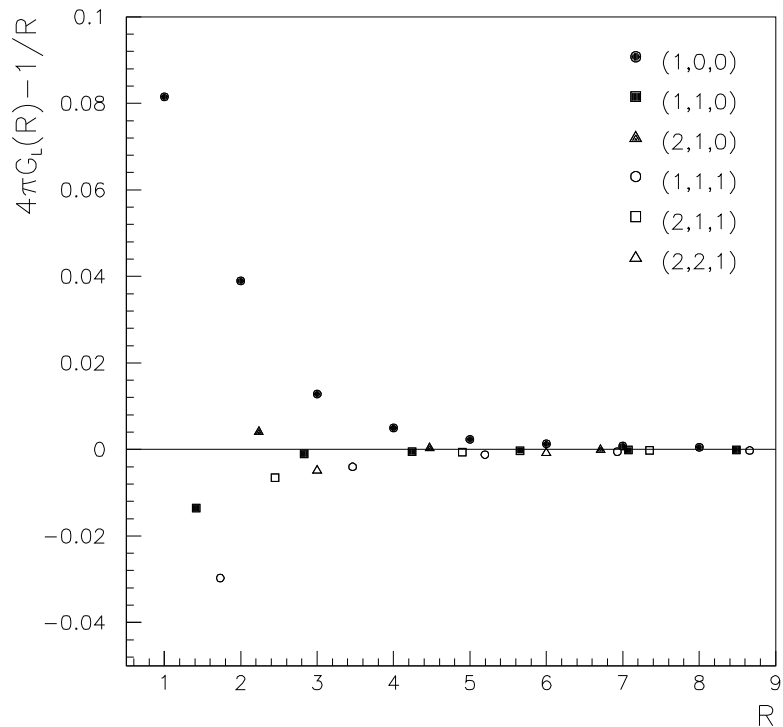}
\end{center}
\caption{\em Difference between the lattice one gluon exchange and the
continuum $1/R$.}
\label{fig5}
\end{figure}

In order to increase the energy resolution we have to use information,
obtained from $q\bar{q}$ separations where effects of
broken continuum $O(3)$ symmetry are still visible. The corresponding values of
the force can be corrected from theoretical
knowledge about the nature of
these violations. The one gluon exchange, which causes a continuum
potential $V_c({\bf R})\propto -1/R$, is replaced by the Fourier
transform of the lattice gluon propagator~\cite{Rebbi}
\begin{equation}
\frac{1}{R}\longrightarrow\left[\frac{1}{\bf R}\right]:=4\pi G_L({\bf R})
=4\pi\int_{-\pi}^{\pi}\!\frac{d^3\!p}{(2\pi)^3}\,
\frac{\cos({\bf Rp})}{4\sum_i\sin^2(p_i/2)}\quad.
\end{equation}
In Fig.~\ref{fig5}, the difference between the two tree-level results for all
$q\bar{q}$ directions, realized in our simulations, is
displayed. As can be seen, from $R=4$ onwards it becomes negligible,
and it is
largest for the on-axis (1,0,0) separations.
The one-loop coefficient has been calculated for the on-axis values
only~\cite{oneloop}, and gives --- apart from a renormalization of the
bare coupling parameter
\begin{equation}
\label{kar}
V({\bf R})\propto -\left[\frac{1}{\bf
R}\right]\frac{1}{\ln(\Lambda_V(R)R)}\quad.
\end{equation}
For large lattices and $R>1$, the deviations of
$\Lambda_V(R)$ from the continuum
$\Lambda_V$ are less than $5\%$, and for the off-axis
directions, they are probably even smaller.

Exploiting this knowledge, we will reconstruct the continuum force
$F(R)$ and check our procedure by comparing results from four
different lattice spacings reaching from $2$~GeV to more than $6$~GeV.
Then, we will define the coupling from the interquark force
\begin{equation}
\label{alpha}
\alpha_{q\bar q}(R)=-\frac{2N}{N^2-1}F(R)R^2\quad.
\end{equation}
Finally, we calculate
$\Lambda_R=1.048\Lambda_{\overline{M\!S}}$~\cite{billoire} from the
running of the coupling according to
\begin{equation}
\label{alpha2}
\alpha_{q\bar q}(R)
        =\frac{1}{4\pi}\left(b_0\ln\left(Ra\Lambda_R\right)^{-2}
        +b_1/b_0\ln\ln\left(Ra\Lambda_R\right)^{-2}\right)^{-1},
\end{equation}
and translate our result into the $\overline{M\!S}$ scheme:
$\alpha_{\overline{M\!S}}(q)=\alpha_{q\bar
q}(r) - 0.082\alpha_{q\bar q}^2(r)+\cdots$ with $r=1/q$.
It is this small two-loop coefficient (0.082), and the dependence on
just a single parameter ($\Lambda_R$),
which turn the interquark
force $F(R)$ into a more user-friendly quantity than the potential
$V(R)$ itself is.

\subsection{Reconstruction of the continuum force}

\begin{figure}[htb]
\begin{center}
\leavevmode
\epsfxsize=400pt
\epsfbox[0 160 600 560]{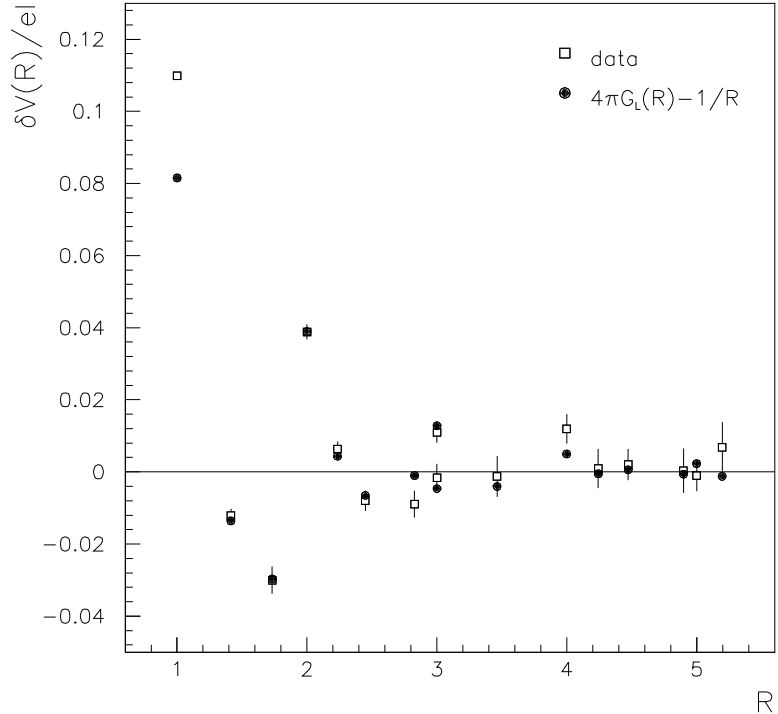}
\end{center}
\caption{\em Comparison between deviations of the potential at $\beta=6.4$
from the continuum symmetry $\delta V({\bf R})/el$,
and $[1/{\bf R}]-1/R$,
which has been used to parametrize lattice artefacts.}
\label{fig6}
\end{figure}

We start from a parametrization of our data on the
$q\bar{q}$-potential which has been invented by Michael~\cite{chris}
\begin{equation}
\label{potfit}
V({\bf R})=V_0+KR-\frac{e}{R}+\frac{f}{R^2}-\delta V({\bf R})\quad,
\quad \delta V({\bf R})=el\left(\left[\frac{1}{\bf R}\right]
-\frac{1}{R}\right)\quad.
\end{equation}
The self energy $V_0$, the string tension $K$, the Coulomb coefficient
$e$, the lattice symmetry correction-coefficient $l$, and $f$ are fit
parameters. $l$ turns out to be $l\approx 0.6$ for $\beta=6.0$,
6.2, 6.4, and 6.8, while $f$ increases from $0.04$ at $\beta=6.0$ up
to $0.10$ at $\beta=6.8$ as it is qualitatively expected
from the running of the
coupling. The $\chi^2$ values have been found to
be acceptable for the fit ranges
$R\geq\sqrt{2}$ ($\beta=6.0$, $\beta=6.2$) and $R\geq\sqrt{3}$
($\beta=6.4$, $\beta=6.8$).
The quality of the fits (for the example $\beta=6.4$) is visualized in
Fig.~\ref{fig6} where
the theoretical value
$\left[1/{\bf R}\right]-1/R$ (circles)
is compared against
$\delta V({\bf R})/(el)$ (squares), as computed from the data.
As a check of the stability of the parametrization we have
additionally performed four parameter fits by constraining $f$ to {\em
zero}. The
parameters $e$ turned out to be smaller in these cases, while
$l\approx 0.6$ was remarkably stable. Only the accepted fit ranges
changed to $R\geq 2$ ($\beta=6.0$, $\beta=6.2$), and $R\geq\sqrt{5}$
($\beta=6.4$, $\beta=6.8$), respectively.

The force can be calculated from the potential by
\begin{equation}
F(\overline{R})=\frac{V({\bf R}_2)-V({\bf R}_1)}{2d}+
{\cal O}\left(\left(2d/R\right)^{\!2}\right)
\end{equation}
with
\begin{equation}
\overline{R}=R\left(1+{\cal O}\left(\left(2d/R\right)^{\!2}\right)\right)
\quad,\quad R=\frac{R_1+R_2}{2}\quad,\quad d=\frac{R_1-R_2}{2}\quad.
\end{equation}
The smaller (larger) $d$ is, the larger the statistical (systematic)
errors become. The systematic errors stem from approximating the
derivative by a finite difference. As a compromise, we have
calculated the force from all pairs of potential values with
distances restricted to $2\geq 2d\geq 1$.
In addition, an assumption about the smoothness of the resulting force
has been made.
As an illustrative example, let us assume
$V(R)=V_0-e/R+KR$ (Eq.~\ref{cornell}). Using the
above prescription with $R_1=R-1$ and
$R_2=R+1$, we obtain $F(R)=-e/(R^2-1)-K$. Since the expression is
correct up to ${\cal O}(4/R^2)$ only, we can substitute $F(R)$ by
$F(\overline{R})$ with
$\overline{R}:=\sqrt{R^2-1}=R\left(1-1/(2R^2)+\cdots\right)$.
This leads to the
correct derivative
$F(\overline{R})=-e/\overline{R}^2-K$.
Exactly the same idea is used to calculate $\overline{R}(R,l,f\!/\!e)$ from
the potential parametrization Eq.~\ref{potfit}.
This was done for $f=f_{\mbox{\scriptsize fit}}$, as well as for $f=0$.
For $l$ a systematic error $\Delta l=l/5$ has been allowed.

\begin{figure}[htb]
\begin{center}
\leavevmode
\epsfxsize=500pt
\epsfbox[25 160 625 560]{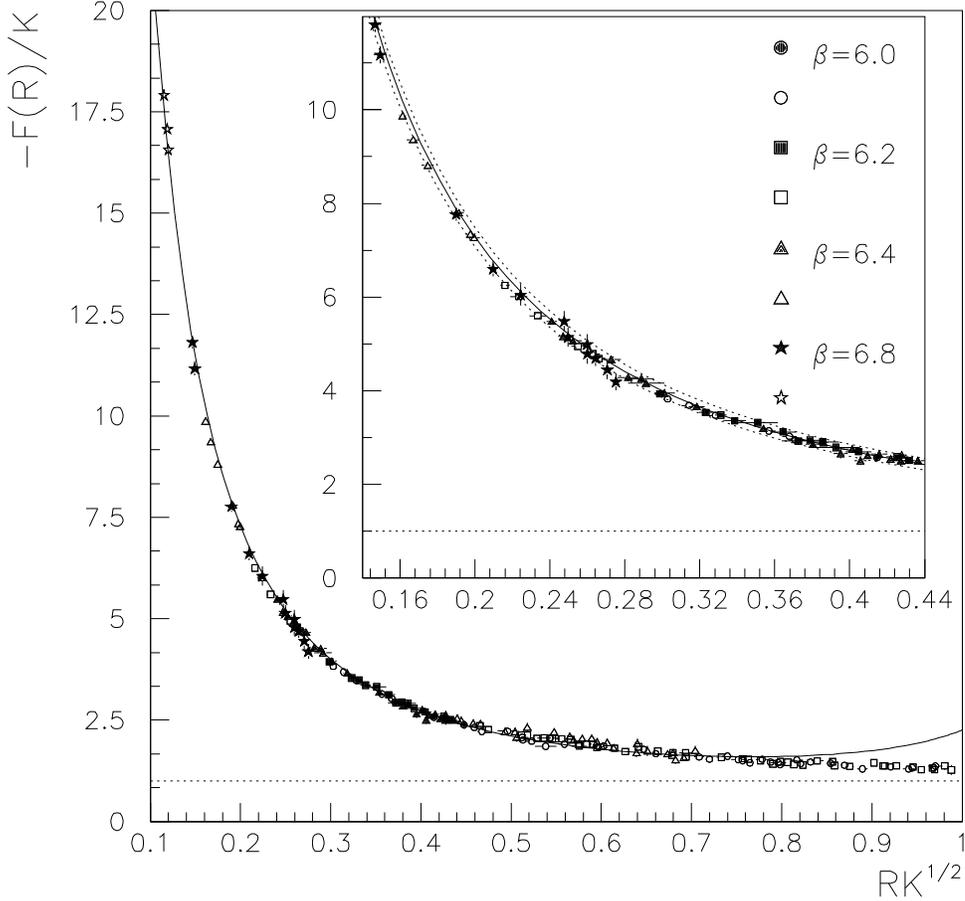}
\end{center}
\caption{\em The reconstructed continuum force from four $\beta$-values is
displayed, together with the curve
$F(R)=-\frac{4}{3}\alpha_{q\bar{q}}(R)/R^2$ for the parameter
$\Lambda_R=0.66(4)\sigma^{1/2}$.
Horizontal errorbars denote systematic errors from
taking the discrete derivative, vertical errorbars are statistical.
The perturbative region is enlarged. Points with open symbols have
been omitted from the determination of $\Lambda_R$.}
\label{fig7}
\end{figure}

In Fig.~\ref{fig7} we display the resulting continuum
force for all our $\beta$-values.
The systematic errors from computing the discrete derivatives are
included as horizontal errorbars while the statistical errors are
shown as vertical errorbars. For small $R$ the systematic uncertainties
dominate. The success of our method is illustrated by the nice scaling
behaviour over the four $\beta$-values down to unexpected small
distances (1.4 lattice units!).

Let me finally emphasize,
that the fit to Eq.~\ref{potfit} has just been performed
to estimate the relative weight of the one-gluon-exchange
$l=0.60\pm 0.12$. $f$ is varied to account for
systematic errors, arising from deviations from a Coulombic behaviour,
due to the logarithmic running of the coupling. No additional information
from the fits has been used for the numerical differentiation of the
potentials.

\begin{table}
\begin{center}
\begin{tabular}{|c|c|c|c|}
\hline
$\beta$ & $R$-range & $a\Lambda_R$ & $\Lambda_R/\sqrt{\sigma}$\\\hline
6.0 & 1.9--2.0 & 0.1440\err{43} & 0.653\err{20}\\
6.2 & 1.9--2.7 & 0.1043\err{25} & 0.659\err{16}\\
6.4 & 2.0--3.7 & 0.0780\err{24} & 0.661\err{21}\\
6.8 & 2.0--7.0 & 0.0458\err{52} & 0.656\err{76}\\
\hline
\end{tabular}
\caption{\em The plateau, in which a constant $\Lambda_R(R)$ has been
observed, and the averaged $\Lambda_R$ in lattice units and in units of
the string tension, respectively.}
\label{tab2}
\end{center}
\end{table}

\begin{figure}[tbh]
\begin{center}
\leavevmode
\epsfxsize=500pt
\epsfbox[0 60 600 350]{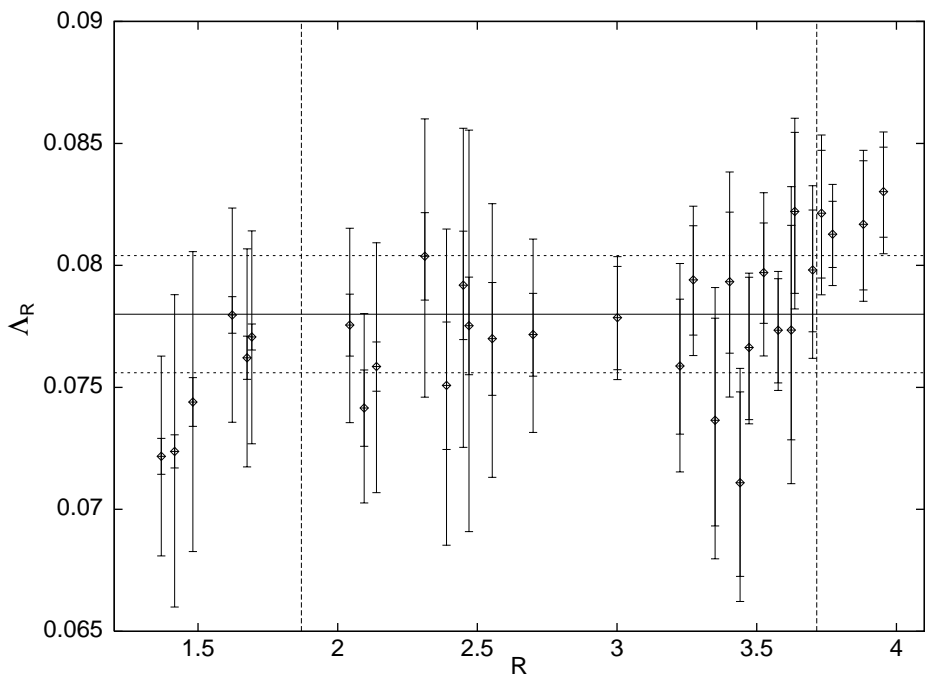}
\end{center}
\caption{\em The plateau in $\Lambda_R(R)$ for $\beta=6.4$.}
\label{fig8}
\end{figure}

\subsection{Calculation of the running coupling}
All data points on the force with $R_1=1$ are excluded from our analysis
to avoid uncontrolled lattice artefacts. For all $\beta$ and $R$ values
$\Lambda_R(R)$ is
calculated by inverting the
relation
\begin{equation}
F(R)=\frac{1}{6\pi b_0}\frac{1}{R^2}\left(\ln(Ra\Lambda_R(R)
)-\frac{b_1}{2b_0}\ln\left(-2\ln(Ra\Lambda_R(R)
)\right)\right)^{-1}\quad,
\end{equation}
obtained by combining Eq.~\ref{alpha} with Eq.~\ref{alpha2}. This
ratio is plotted against $R$. A plateau is identified for $Ra<
1$~GeV$^{-1}$ in all cases. As an example we show the $\beta=6.4$ data
in Fig.~\ref{fig8}. The smaller errorbars indicate the statistical
errors. The systematic errors are included into the larger errorbars.
Below $(Ra)^{-1}\approx 1$~GeV the
$\Lambda_R(R)$ values increase a bit, and below about $0.5$~GeV the
values decrease towards zero. Since the data on the force is
correlated, and systematic and statistical errors cannot clearly be
disentangled, we do not attempt to fit the force but average
the $\Lambda_R(R)$ values over the plateau.
The results are shown in Tab.~\ref{tab2}. At $\beta=6.4$, where
$\Lambda_R(R)$ remains constant over a change of a factor {\em two}
in the energy scale,
and at $\beta=6.8$, where this factor is about {\em four}, these plateaus
can clearly be identified. The 6.0 and 6.2 data yield
consistent but less compelling results. So,
we do not attempt to use them in our final
analysis. Averaging the $\beta=6.4$ and $\beta=6.8$ results, and
allowing an additional $5\%$ systematic error on $\Lambda$,
mainly caused by the one-loop lattice artefacts (Eq.~\ref{kar}),
leads to
\begin{equation}
\Lambda_R=(0.660\pm 0.40)\sqrt{\sigma}\quad.
\end{equation}
The corresponding fit curve is plotted in
Fig.~\ref{fig7}. As can be seen, the data is well described, even for
data points that had been omitted from the $\Lambda$-determination, in
order to exclude discretization errors.
On the other end, the data is in qualitative
agreement with the perturbative prediction down to $0.5$~GeV,
surprisingly close to the pole $\Lambda_R$.

\begin{figure}[htb]
\begin{center}
\leavevmode
\epsfxsize=400pt
\epsfbox[0 160 600 560]{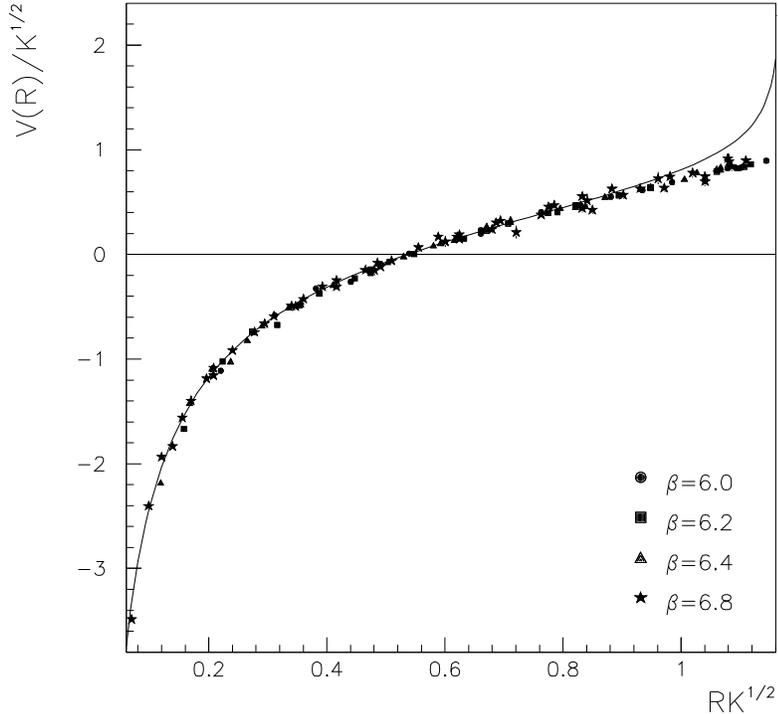}
\end{center}
\caption{\em The uncorrected lattice potential in units
of the string tension, together with
the integrated perturbative
force (solid curve) for $\Lambda_R=0.66\sigma^{1/2}$.}
\label{fig9}
\end{figure}

The above value can be converted into
\begin{equation}
\label{msres}
\Lambda_{\overline{M\!S}}^{(0)}=(0.630\pm 0.38)\sqrt{\sigma}
=(293\pm 18^{+25}_{-63}) \mbox{MeV}\quad,
\end{equation}
using the scale estimate Eq.~\ref{sca}. The second error is the
overall scale error from the experimental uncertainty in
$\sqrt{\sigma}$. The four $\Lambda_{\overline{M\!S}}$ values, extracted
from the force at the different $\beta$-values, as well as our final,
averaged
result (dashed errorband) are displayed in Fig.~\ref{fig4}. The values scale
so well that one might wonder whether we have overestimated the
systematic errors. This question cannot be answered unless
one-loop perturbative results on the off-axis potential are available.
In Fig.~\ref{fig9}, a comparison of the integrated perturbative force
($\Lambda_R=0.66\sqrt{\sigma}$) with the original, uncorrected
potential is made. As can be seen, the data is described quite well,
even without any lattice artefact corrections. Apparently, the
theoretical curve has not been biased through the analysis procedure.

\subsection{Inclusion of dynamical quark flavours}

How can we relate our result, $\alpha_{\overline{M\!S}}$
for the case of zero fermion flavours,
to the real world that includes sea quarks?
It is shown that our method works in
principle and can be applied to dynamical configurations.
In the meantime, it is possible to estimate the impact of dynamical
quark flavours on the coupling by exploiting other lattice results.
Pioneering studies of full QCD have revealed that the sea quarks lead
to a decrease of the lattice spacing $a$ at fixed $\beta$.
However, dimensionless ratios of physical observables are in qualitative
agreement with quenched results. The shift in $\beta$, necessary to
match a quenched lattice spacing $a$ to the unquenched $a$, is called
$\beta$-shift. In the following, this $\beta$-shift will be utilized
to predict the corresponding change in the running coupling. From this
matching, a large uncertainty in the energy scale arises but, since
the coupling is running logarithmically, this has only a limited
impact on the coupling at high energies, e.g.\ at the
$Z^0$ mass $m_Z$.

\begin{table}
\begin{center}
\begin{tabular}{|c|c|c|c|c|c|c|}
\hline
$\beta^{(4)}$ & $\beta^{(0)}$ & $S_{\Box}^{(4)}$ & $S_{\Box}^{(0)}$&
${g_{\mbox{\scriptsize FNAL}}^{(4)}}^{\!\!\!2}$&
${g_{\mbox{\scriptsize FNAL}}^{(0)}}^{\!\!\!2}$&
$\pi/a\sqrt{\sigma}$\\
\hline
5.15&5.74(9)&0.4760(1)&0.443&2.0597(4)&1.79(3)&$8.5^{+2.0}_{-1.6}$\\
5.35&5.90(3)&0.4470(1)&0.418&1.8910(4)&1.676(9)&$12.5^{+0.5}_{-0.9}$\\
\hline
\end{tabular}
\caption{\em Matching between four flavour QCD and the quenched
approximation by use of the interquark potential. The data in the
first three columns is taken
from Ref.~\protect\cite{MTC}. In the last column, the matching scale has
been calculated.}
\label{tab3}
\end{center}
\end{table}

For our extrapolation of $\alpha_{\overline{M\!S}}$
to $m_Z$, we use results, obtained by the $MT_C$
collaboration~\cite{MTC}, for four flavours of staggered
quarks with a physical mass of about $35$~MeV. In this reference,
the $\beta$-shift has
been computed by matching the dynamical $q\bar{q}$
potential to the static potential.
The results are collected in Tab.~\ref{tab3}. In column~{\em two}
the corresponding
quenched $\beta$ values are displayed. In columns~{\em three} and {\em
four} the plaquette values are collected. In columns {\em five}
and {\em six} $g^2$ has been
calculated in the FNAL-$\overline{M\!S}$-scheme.
In the last column, the matching momenta $\pi/a$ in units of the
string tension from our analysis are displayed.
They are comfortably within the perturbative region.
A matching between the four fermion running coupling and the
quenched coupling is done, yielding
$\Lambda_{\overline{M\!S}}^{(4)}=(0.44\pm
0.10)\Lambda_{\overline{M\!S}}^{(0)}$.
The scale $\pi/a$ cancels from this ratio.
The error consists of a $12\%$ part from the matching of both
couplings at different $\pi/a$, and a further
$10\%$ from possible scaling
violations\footnote{As can be seen from Fig.~\ref{fig4},
$\sqrt{\sigma}/\Lambda_{\overline{M\!S}}$,
computed in the
FNAL-$\overline{MS}$-scheme,
still differs by about
$25\%$ from its asymptotic value at
$a>0.1$~fm~$\approx 0.25\sigma^{-1/2}$.}, which might
behave differently in the case of dynamical
flavours (compared to the quenched case).

Real nature has non-degenerate quark flavours. Thus, we have to incorporate
systematic errors emanating from the use of $m_u=m_d=m_s=m_c=35$~MeV
into our prediction. What is the effect of shifting $m_c$ up to
1~GeV? If we evolve the coupling from 1~GeV down to 0.5~GeV, an
energy, where perturbation theory at least qualitatively describes the
running of $\alpha_{q\bar{q}}$, by use of the four flavour
$\beta$-function, evolve it back with the three flavour
$\beta$-function, and match it to the four flavour formula again (at
1~GeV), we
find a decrease of $\Lambda_{\overline{M\!S}}^{(0)}$ by $10\%$.
A shift into the opposite direction of approximately the same
magnitude~\cite{hase} is caused by the fact that the light quarks have been
too heavy on the lattice. So, we do not intend to change
our central value but increase the systematic error and end up with
\begin{equation}
\Lambda_{\overline{M\!S}}^{(4)}=0.44^{+10}_{-14}
\Lambda_{\overline{M\!S}}^{(0)}=(129\pm 8^{+43}_{-60}) \mbox{MeV}\quad.
\end{equation}
This can be converted into a value for five flavours with an
additional error from matching the couplings at $0.75m_b<\mu<2m_b$ with
$m_b=4.8$~GeV. The final result for five active quark flavours is
\begin{equation}
\alpha_{\overline{M\!S}}\left({m_Z}^{+45\%}_{-60\%}\right)=0.102
\end{equation}
or, converting the scale error into an error on
$\alpha_{\overline{MS}}$, this value is
\begin{equation}
\alpha_{\overline{M\!S}}\left(m_Z\right)=0.1020^{+054}_{-104}\quad.
\end{equation}
Even, if we assume the extreme case
$\Lambda_{\overline{M\!S}}^{(4)}=\Lambda_{\overline{M\!S}}^{(0)}$, it is
impossible to obtain values $\alpha_{\overline{M\!S}}(m_Z)>0.115$ from
the lattice.
However, the effects of dynamical fermions, as well as the exact influence
of quark masses~\cite{shirkov}, have to be investigated
carefully in the future.

\section{Summary and Outlook}

With a medium scale supercomputer simulation of the valence quark
approximation of QCD, a running of the strong
coupling ``constant'', as predicted by perturbation theory, has been
found in the region between 1~GeV and 5~GeV. Qualitative agreement with a
two-loop running was observed down to energies of about $0.5$~GeV. From
this, the QCD $\Lambda$-parameter (or
alternatively the coupling at a given high energy scale) can be
related to low energy observables like the hadron spectrum.
Since the renormalization scheme in which the coupling was
``measured'', is very close to the $\overline{M\!S}$-scheme, it is
likely that the latter scheme gives reliable results
down to energies as low as
1~GeV. Alternatively, one can use the lattice to define a
renormalization scheme in a nonperturbative gauge-invariant fashion,
e.g.~the coupling from the interquark force.

\begin{table}
\begin{center}
\begin{tabular}{|c|c|c|c|}
\hline
Source of uncertainty&present error&future error&solution\\
\hline
$\Lambda_R$&$3\%$&$<3\%$&smaller $a$, higher statistics\\
$\Lambda_R=\Lambda_R(R)$&$5\%$&$0\%$&one-loop off-axis potential\\
$\sqrt{\sigma}=$~?&$15\%$&$<5\%$&dynamical fermions\\
$\Lambda_{\overline{M\!S}}^{(0)}\rightarrow\Lambda_{\overline{M\!S}}^{(4)}$&
$25\%$&$<5\%$&dynamical fermions\\
$\Lambda_{\overline{M\!S}}^{(4)}\rightarrow\Lambda_{\overline{M\!S}}^{(5)}$&
$2\%$&$2\%$&---\\\hline
\end{tabular}
\caption{\em Error sources in the calculation of
$\Lambda_{\overline{M\!S}}^{(5)}$, and how these uncertainties can be
removed/reduced.}
\label{tab4}
\end{center}
\end{table}

Combining the result with phenomenological potential
models~\cite{elkhadra} or data from lattice simulations of full QCD, a
real world value can be estimated. The systematic part of the error, mainly
caused by neglecting quark loops in the simulation, is dominant.
An upper limit for $\alpha_{\overline{M\!S}}(m_Z)$ is found:
$\alpha_{\overline{M\!S}}(m_Z)<0.115$.
In Tab.~\ref{tab4}, sources of the scale errors, and how they
might be removed in the near future, are collected. It should be
notet, however, that, despite all
systematic uncertainties, the accuracy of lattice simulations can
compete with present day experiments.

The method can immediately be applied to full (lattice) QCD.
The computational effort of the
present investigation has been moderate, compared to most other lattice
studies of quantities of direct experimental relevance.
So, an analoguous computation of the $q\bar{q}$ potential with
dynamical fermions will become feasible
in the near future. It should be amongst the earliest experiments
on the first generation of Tera\-FLOPS computers. It will be very
interesting to study the dependence of the running of $\alpha_S$ on
the opening of new quark flavours. The impact of the quark masses
and the transition behaviour at flavour thresholds can also be
investigated.
\eject

\noindent{\large\bf Acknowledgements}

\noindent
I would like to thank Klaus Schilling for carefully proofreading the
manuscript, and for his helpful comments and suggestions.
Edwin Laermann is thanked for
discussions and for letting me know the plaquette values of the $MT_C$
simulations. The computations have been performed on the
8K Connection Machine CM-2
of the Institut f\"ur Angewandte Informatik, Wuppertal. I am grateful
to Peer Ueberholz and Randy
Flesch for their commitment to the Wuppertal CM project
and for constant help and advice.
The CM project has been supported by the DFG (grant Schi 257/1-4).
This research has been funded by the EC grant \#~SC1*-CT91-0642.

\vfill\eject


\begin{thebibliography}{99}

\bibitem{weingarten} {F.~Butler, H.~Chen, J.~Sexton, A.~Vaccarino,
and D.~Weingarten, {\em Phys.~Rev.~Lett.}\ {\bf 70} (1993) 2849.}

\bibitem{wir} {G.S.~Bali and K.~Schilling, {\em Phys.~Rev.}\
   {\bf D46} (1992) 2636.}

\bibitem{peran} {C.~Michael and S.~Perantonis, {\em Nucl.~Phys.}\ {\bf
  B347} (1990) 864.}

\bibitem{wir2} {G.S.~Bali and K.~Schilling, {\em Phys.~Rev.}\
   {\bf D47} (1993) 661.}

\bibitem{wir3} {K.~Schilling and G.S.~Bali, Wuppertal preprint WUB
93-33 (1993), talk by K.~Schilling at the Workshop on Large Scale
Computational Physics, HLRZ J\"ulich.}

\bibitem{wir4} {G.S.~Bali and K.~Schilling, in preparation.}

\bibitem{eicht1} {E.~Eichten and K.~Gottfried, {\em Phys.~Lett.}\ {\bf
66B} (1977) 286.}

\bibitem{quigg}{C.~Quigg and J.L.~Rosner, {\em Phys.~Rept.} {\bf 56C}
(1979) 167.}

\bibitem{eicht2} {E.~Eichten, K.~Gottfried, T.~Kinoshita, K.D.~Lane,
and T.M.~Yan, {\em Phys.~Rev.} {\bf D21} (1980) 203.}

\bibitem{sommer} {R.~Sommer, DESY preprint DESY-93-062 (1993).}

\bibitem{ukqcd} {UKQCD Collaboration: G.S.~Bali,
A.~Hulsebos, A.C.~Irving, C.~Michael,
P.W.~Stephenson, and K.~Schilling, {\em Phys.~Lett.} {\bf B309}
(1993) 378.}

\bibitem{mack1} {G.P.~Lepage and P.B.~Mackenzie,
{\em Nucl.~Phys.}\ {\bf B}[Proc.~Suppl.]{\bf 20} (1991) 173.}


\bibitem{parisi} {G.~Parisi, Proceedings of the {\sc xx}th
International Conference on High Energy Physics 1980, Madison, Eds.\
L.~Durand, and L.G.~Pondrom, American Institute of Physics, New York
(1981) 1531.}

\bibitem{mack} {A.X.~El-Khadra, G.~Hockney, A.S.~Kronfeld, and
P.B.~Mackenzie, {\em Phys.~Rev.~Lett.}\ {\bf 69} (1992) 729;
P.B.~Mackenzie,
{\em Nucl.~Phys.}\ {\bf B}[Proc.~Suppl.]{\bf 26} (1992) 369.}

\bibitem{mack2} {G.P.~Lepage and P.B.~Mackenzie, {\em Phys.~Rev.}\ {\bf
D48} (1993) 2250.}

\bibitem{msbar}{R.~Dashen and D.J.~Gross, {\em Phys.~Rev.}\
               {\bf D23} (1981) 2340.}

\bibitem{Hasenfratz} {A.~and P.~Hasenfratz, {\em Phys.~Lett.}\
                 {\bf 93B} (1980) 165;
                  {\em Nucl.~Phys.}\ {\bf B193} (1981) 210.}

\bibitem{digiac} {A.~DiGiacomo and G.C.~Rossi, {\em Phys.~Lett.}\
                 {\bf 100B} (1981) 481;
                  A.~DiGiacomo and G.~Paffati, {\em Phys.~Lett.}\
                 {\bf 108B} (1982) 327.}

\bibitem{oneloop} {U.~Heller and F.~Karsch, {\em Nucl.~Phys.}
                 {\bf B251} (1985) 254.}

\bibitem{pana} {B.~Alles, M.~Campostrini, A.~Feo, and H.~Panagopoulos,
PISA preprint IFUP-TH-17-93 (1993).}

\bibitem{karsch} {F.~Karsch and R.~Petronzio, {\em Phys.~Lett.}\ {\bf 153B}
(1985) 87.}

\bibitem{wir1} {G.S.~Bali and K.~Schilling, {\em Nucl.~Phys.}\
{\bf B}[Proc.~Suppl.]{\bf 30} (1993) 513.}

\bibitem{elkhadra} {A.X.~El-Khadra, G.M.~Hockney, A.S.~Kronfeld,
and P.B.~Mackenzie, Proceedings of the 26th International Conference on
High Energy Physics, Dallas (1992) 1523; A.X.~El-Khadra, talk presented at
the Lattice '93 conference, Dallas.}

\bibitem{luesch} {M.~L\"uscher, R.~Narayanan, P.~Weisz, and U.~Wolff,
{\em Nucl.~Phys.}\ {\bf B384} (1992) 168.}

\bibitem{luescher} {M.~L\"uscher, R.~Sommer, U.~Wolff, and P.~Weisz,
{\em Nucl.~Phys.}\ {\bf B389} (1993) 247; DESY preprint DESY-93-114
(1993).}

\bibitem{chris} {C.~Michael, {\em Phys.~Lett.}\ {\bf B283} (1992) 103;
UKQCD Collaboration: S.P.~Booth, K.C.~Bowler,
D.S.~Henty, R.D.~Kenway, B.J.~Pendleton,
D.G.~Richards, A.D.~Simpson, A.C.~Irving, A.~McKerrell, C.~Michael,
P.W.~Stephenson, M.~Teper, and K.~Decker, {\em Phys.~Lett.}\
{\bf B275} (1992) 424; UKQCD Collaboration: S.P.~Booth, D.S.~Henty,
A.~Hulsebos, A.C.~Irving, C.~Michael, and P.W.~Stephenson,
{\em Phys.~Lett.}\ {\bf B294} (1992) 385.}

\bibitem{Rebbi} {C.B.~Lang and C.~Rebbi, {\em Phys.~Lett.}\
                      {\bf 115B} (1982)137.}

\bibitem{billoire} {A.~Billoire, {\em Phys.~Lett.}\ {\bf 104B} (1981)
                    472.}

\bibitem{MTC} {$MT_C$ Collaboration: K.D.~Born, R.~Altmeyer,
               W.~Ibes, E.~Laermann, R.~Sommer,
               T.F.~Walsh, and P.~Zerwas,
               {\em Nucl.~Phys.}\ {\bf B}[Proc.~Suppl.]{\bf 20} (1991)
               394; E.~Laermann, private communication.}

\bibitem{hase} {A.~Hasenfratz, T.A.~DeGrand, Colorado preprint
COLO-HEP-311 (1993).}

\bibitem{shirkov} {D.V.~Shirkov, {\em Theor.~Math.~Phys.}\ {\bf 93}
(1992) 1403.}

\end{thebibliography}
\end{document}